\documentclass[letterpaper,aps,superscriptaddress,showpacs,floatfix,twocolumn,10pt]{revtex4-1}
\usepackage{graphicx}
\usepackage{amsmath}
\usepackage{graphicx}

\begin{document}
\title{Chemical properties of super-hadronic matter created in relativistic heavy ion collisions}
\author{Scott Pratt}
\author{William Patrick McCormack}
\affiliation{Department of Physics and Astronomy and National Superconducting Cyclotron Laboratory\\
Michigan State University, East Lansing, MI 48824~~USA}
\author{Claudia Ratti}
\affiliation{Universit\'a degli Studi di Torino and INFN Torino, via
Giuria 1-10125, Torino~~ITALY}
\affiliation{Department of Physics, University of Houston, Houston, TX
77204, USA}
\date{\today}


\begin{abstract}
Preliminary charge balance functions from the STAR Collaboration at the Relativistic Heavy Ion Collider (RHIC) are compared to a model where quarks are produced in two waves. If a chemically equilibrated quark-gluon plasma (QGP) is created the strength and diffusive spread of the first wave should be governed by the chemical composition of the QGP, while the second wave should be determined by the increased number of quarks required to make the observed final-state hadrons. A simple model parameterizes the chemistry of the super-hadronic matter and the two correlation lengths for the two waves. Calculations are compared to preliminary data from the STAR Collaboration. The chemistry of the super-hadronic matter appears to be within 20\% of expectations from lattice gauge theory.
\end{abstract}

\maketitle

\section{Introduction}
The quark-gluon plasma (QGP) has a remarkable chemistry. Counting spins, colors and flavors, there are 36 light degrees of freedom from the up, down and strange quarks, and an additional 16 from gluons. Thus, approximately 52 particles should inhabit a volume on the order of one thermal wavelength cubed, $\sim (\hbar c/T)^3$. The matter created in high-energy heavy ion collisions at RHIC and at the LHC achieves energy densities on the scale of 10 GeV/fm$^3$, well above the threshold for defining individual hadrons. However, the chemical properties of the super-hadronic matter have not yet been shown to match expectations of a chemically equilibrated QGP. Here we report on how charge balance functions measured by STAR make a strong case that the chemical makeup of the matter produced in heavy ion collisions is within a few tens of percent of that of the QGP. Further, it appears that charge production comes in two waves, the first wave being the creation of the QGP, and the second being hadronization.

For a non-interacting gas of massless partons, the density of each parton species scales with the entropy density. Thus, in an isentropic expansion of a weakly interacting parton gas, where one expects the net entropy within a given rapidity window to be conserved, the number of each species within that same window would also remain constant until the system begins to hadronize. In order to conserve entropy during hadronization, a large number of quark-antiquark pairs must be produced, which was predicted to lead to narrower separation for balancing charges in central collisions \cite{Bass:2000az}, an effect observed by STAR \cite{Adams:2003kg}. Thus quark-antiquark pairs would mainly be created in two waves, the first being when the QGP is formed and the second at hadronization. Quark-antiquark pairs created in the first wave might significantly separate in coordinate space, perhaps on the scale of a unit of spatial rapidity, $\Delta\eta =(1/2)\Delta\ln(t+z)/(t-z)$. In contrast, pairs created during the second wave might separate by only a few tenths of a unit of spatial rapidity due to the limited time for separation. Charge balance correlations are built on like-sign subtractions and are thus mainly driven by the effects of charge conservation. In \cite{Pratt:2012dz} ``generalized'' charge balance functions were proposed as a more discriminating measure by considering the correlation of any two species, e.g.,
\begin{equation}\label{eq:Galphabeta}
G_{\alpha\beta}(\Delta\eta)=
\langle[N_\alpha(\Delta\eta)-N_{\bar{\alpha}}(\Delta\eta)]
[N_\beta(0)-N_{\bar{\beta}}(0)]\rangle.
\end{equation}
Here $\alpha$ or $\beta$ refer to specific hadron species, e.g. $\pi^+$, and $\bar{\alpha}$ refers to its antiparticle. For this paper, boost invariance, equivalently translational invariance along the beam axis, is assumed so that one can consider $G(\Delta\eta)$ rather than the more general $G(\eta+\Delta\eta,\eta)$. In the simple model considered here, each balance function will have two contributions, corresponding to the two waves of charge production. The strength of the two contributions are determined by the final state yields and by the chemistry of the super-hadronic matter. If the super-hadronic matter is a chemically equilibrated QGP, most of the electric charge is created in the second wave, whereas most of the strangeness is created in the first wave. Because most of the electric charge is in pions one expects $G_{\pi^+\pi^-}(\Delta\eta)$ to have a width determined by how far balancing charges separate given they were produced toward the end of the evolution. In contrast, most of the strangeness is carried by kaons and the width of the ${K^+K^-}$ balance function should be broader and reflect how far those charge pairs created early in the reaction might separate. One would expect the $p\bar{p}$ balance function to be largely determined by baryon conservation. Even though many quark-antiquark pairs are produced at hadronization, the baryon number fluctuation defined below will fall if the number of final-state baryons is less than one ninth of the number of quarks before hadronization. The contribution to $B_{p\bar{p}}$ from the second wave could then be negative, leading to a dip in the balance function at small $\Delta\eta$. These heuristic arguments rely on a simple picture of the super-hadronic chemistry, and neglect the fact that hadrons carry multiple charges, e.g. a $K^+$ carries both strangeness and electric charge. These shortcomings will be addressed below.
\begin{figure}
\centerline{\includegraphics[width=0.35\textwidth]{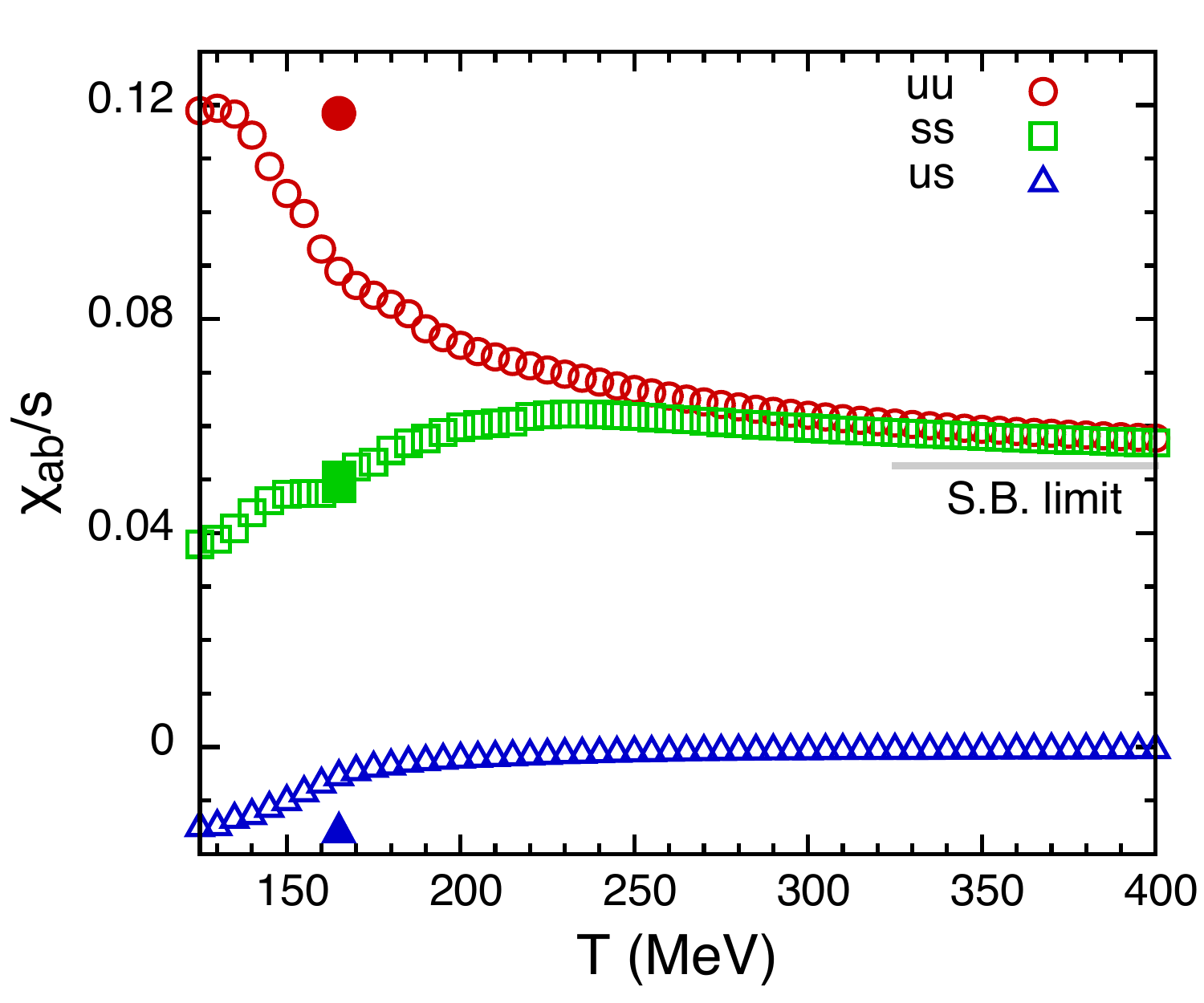}}
\caption{\label{fig:claudia}
Charge fluctuations from lattice gauge theory \cite{Borsanyi:2011sw,Ratti:2011au} (open symbols) are similar to those of a hadronic gas (filled symbols for $T=165$ MeV). For fixed entropy there are increased numbers of up and down quarks in the hadronic phase, whereas the number of strange quarks is slightly smaller. The off-diagonal element disappears above $T_c$ when hadrons dissolve and quark-antiquark correlations disappear. At high temperatures the results approach those of a Stefan-Boltzmann gas of massless partons (S.B. limit).
}
\end{figure}

The assertion that the ratio of quarks to entropy stays roughly constant above the transition temperature has been confirmed by lattice calculations of charge fluctuations \cite{Borsanyi:2011sw,Ratti:2011au} as shown in Fig. \ref{fig:claudia}. The charge fluctuation for a system with no net charge, $\langle Q_a\rangle=0$, is defined
\begin{equation}
\chi_{ab}\equiv\langle Q_aQ_b\rangle/V,
\end{equation}
Here, $Q_a$ is the net up, down or strange charge within a volume $V$. Assuming zero net charge is reasonable for mid-rapidity measurements at the LHC or at the highest RHIC energies. The fluctuations, $\chi_{ab}$, are well-defined observables that serve as a proxy for quark densities. In a parton gas $\chi_{ab}^{\rm(qgp)}=\delta_{ab}(n_a+n_{\bar{a}})$, where $n_a$ and $n_{\bar{a}}$ are the densities of up and down quarks. This follows because for a non-interacting gas, the only correlations are those between a particle and itself and for a hadron gas $\chi_{ab}^{\rm(had)}=\sum_\alpha n_\alpha q_{\alpha,a}q_{\alpha,b}$, where $q_{\alpha,a}$ is the charge of type $a$ of a hadron where the species is labeled by $\alpha$. The approximate constancy of $\chi_{ab}/s$ in Fig. \ref{fig:claudia} suggests that the number of quarks remains roughly constant within an expanding and co-moving hydrodynamic volume element. The rise of the ratio at low temperature is driven by the rise of the total number of quarks in the hadronic stage. Further, the off-diagonal element $\chi_{us}$ can come from two quarks being in the same hadron, or by hadron-like correlations in the QGP \cite{Koch:2005vg}. 

The purpose of this letter is to show how recent STAR measurements of balance functions can be used to validate the two-wave nature of charge formation, and to extract the ratio $\chi_{ab}/s$ in the QGP phase. To do this, we must overcome the obvious constraint that, unlike lattice calculations charge does not fluctuate when integrated over the entire collision volume. Further, one must look through the haze of hadronization. This will be accomplished by focusing on charge correlations, which describe the probability for having two charges separated in spatial rapidity by $\Delta\eta$,
\begin{equation}
g_{ab}(\Delta\eta)=\langle \rho_a(0)\rho_b(\Delta\eta)\rangle,
\end{equation}
where $\rho_a$ is the charge density of type $a$, and could be a density per unit volume or per unit spatial rapidity, depending on the context. Even though $\langle\rho_a\rangle=0$, the correlation $g_{ab}$ can differ from zero due to correlations. Here we review how $g_{ab}$ can be used to extract $\chi_{ab}$.

Charge conservation demands $\int d\Delta\eta~g_{ab}(\Delta\eta)=0.$ Assuming that $\chi_{ab}$ in the grand-canonical treatment is driven solely by short range correlations, one can state that the correlation for times before hadronization should have a form similar to
\begin{equation}
g_{ab}(\Delta\eta)=\chi_{ab}^{\rm(qgp)}\left\{
\delta(\Delta\eta)-\frac{e^{-(\Delta\eta)^2/2\sigma^2_{\rm (qgp)}}}{(2\pi\sigma^2_{\rm(qgp)})}
\right\}.
\end{equation}
More generally, the Gaussian form could be replaced by any function that integrates to unity, but one would expect a Gaussian form if quarks appeared suddenly, at a single time, followed by diffusion \cite{Pratt:2012dz}. Quark-antiquark pairs created by the breaking of longitudinal flux tubes could be pulled apart by the tunneling process. It would not be surprising if the true form for the correlation would be rather non-Gaussian

Immediately after hadronization, the correlation of a particle with itself is defined by $\chi^{\rm(had)}$. The long-range correlation cannot change suddenly, because the newly created pairs are uncorrelated with charge far away. Thus to satisfy the condition that $g_{ab}$ integrates to zero, the correlation could have a form similar to
\begin{eqnarray}\label{eq:gabhad}
g_{ab}(\Delta\eta)&=&\chi_{ab}^{\rm(qgp)}\frac{e^{-(\Delta\eta)^2/2\sigma^2_{\rm (qgp)}}}{(2\pi\sigma^2_{\rm(qgp)})}\\
\nonumber
&&\hspace*{-48pt}
+\chi_{ab}^{\rm(had)}\delta(\Delta\eta)
-(\chi_{ab}^{\rm(had)}-\chi_{ab}^{\rm(qgp)})\frac{e^{-(\Delta\eta)^2/2\sigma^2_{\rm (had)}}}{(2\pi\sigma^2_{\rm(had)})}.
\end{eqnarray}
Here, the width $\sigma_{\rm(had)}$ would be narrow, and in the limit of sudden hadronization one could justify the Gaussian form if the spread were diffusive. The width $\sigma_{\rm(qgp)}$ should be significantly larger and would be determined by how far two balancing charges might separate if they were created during the formation of the QGP.

If one measured all hadrons immediately after hadronization, $g_{ab}$ could be reconstructed from the correlations between hadrons. Unfortunately, neutrons are rarely measured, and weak decays make it impossible to fully reconstruct the strangeness. Thus, we consider correlations of hadrons as defined in Eq. (\ref{eq:Galphabeta}),
In reference \cite{Pratt:2012dz} it was shown that $G_{\alpha\beta}(\Delta\eta)$ could be extracted from $g_{ab}$ by assuming additional charge is spread amongst the various hadrons thermally. The expression is
\begin{eqnarray}
G_{\alpha\beta}(\Delta\eta)&=&4\sum_{abcd}\langle n_{\alpha}\rangle q_{\alpha,a}\\
\nonumber
&&\cdot\chi^{{\rm(had)}(-1)}_{ab}(0)g^{{\rm(had)}}_{bc}(\eta)
\chi^{{\rm(had)}(-1)}_{cd}(\eta)
q_{\beta,d}\langle n_{\beta}\rangle.
\end{eqnarray}
Here $g_{bc}$ comes from Eq. (\ref{eq:gabhad}) and $\chi_{\rm(had)}$ is determined by final state yields. Finally, one must account for the thermal smearing in mapping from spatial rapidity, $\eta$, to momentum rapidity, $y$. As explained in \cite{Pratt:2012dz}, this can be done by convoluting the correlations $G_{\alpha\beta}$ onto a blast wave model where two parameters come into play: the final-state temperature and radial collective flow velocity. These parameters are taken to fit the mean $p_t$ of pions and protons reported by STAR. The collective velocity is $u_\perp=0.732c$, and the dissolution temperature is 102 MeV. Additionally, the decays of emitted particles are also taken into account by mapping the pre-decay correlations onto post-decay correlations by simulating decays. 
Hadron multiplicities are generated by assuming that particles were created according to chemical equilibrium with a temperature of 165 MeV. Baryon annihilations were not included in the thermal yields, but one would expect annihilations to reduce the baryon yields by approximately one third \cite{Pan:2012ne,Steinheimer:2012rd}, an expectation that has some experimental support {Adler:2003cb}. Baryon yields were therefore reduced by 30\%, which significantly affects the $p\bar{p}$ balance function shown below. Calculations were repeated for a lower chemical freeze-out temperature, 150 MeV, but with no reduction in the baryon yield. This choice correctly reproduces experimental fluctuations of conserved charges, and can also reproduce the yields \cite{Borsanyi:2014ewa,Bluhm:2014wha}; the results were nearly indistinguishable from those shown here. Finally, correlations were superimposed onto a model of the STAR acceptance and efficiency. The four adjustable model parameters are summarized in Table \ref{table:parameters}.
\begingroup
\squeezetable
\begin{widetext}
\begin{table*}
\caption{\label{table:parameters}
Model parameters varied in fitting experimental data.}
\begin{tabular}{|c|c|c|c|}
\hline
parameter & description & expectation & MCMC range\\
\hline
$\sigma_{\rm(qgp)}$ & \parbox{4.0in}{\baselineskip=12pt
\vspace*{3pt}Spread of balancing charges created in initial thermalization of QGP\vspace*{3pt}}
& $\sim 1$ & $0.3 - 1.5$\\
$\sigma_{\rm(had)}/\sigma_{\rm(qgp)}$ &\parbox{4.0in}{\baselineskip=12pt
\vspace*{3pt}
Spread of balancing charges created during or after hadronization/$\sigma_{\rm(qgp)}$\vspace*{3pt}} & $\sim 0.25$ & $0 - 1$\\
$(\chi_{uu}+\chi_{dd}+\chi_{ss})/s$ & \parbox{4.0in}{\baselineskip=12pt
\vspace*{3pt}
quarks to entropy in QGP\vspace*{3pt}} & $\sim 0.18$ & 0.05 - 0.35\\
$\chi_{ss}/\chi_{uu}$ &\parbox{4.0in}{\baselineskip=12pt
\vspace*{3pt}
Ratio of strange quarks to up or down quarks\vspace*{3pt}}& $\sim 0.95$ & 0 - 1.3\\
\hline
\end{tabular}
\end{table*}
\end{widetext}
\endgroup

Calculations were compared to preliminary balance functions from STAR \cite{Wang:2012jua}, $B_{\pi^+\pi^-}, B_{K^+K^-}$ and $B_{p\bar{p}}$. Balance functions, $B_{\alpha\beta}$, are related to the correlations defined in Eq. (\ref{eq:Galphabeta}) by a factor of the multiplicity, $B_{\alpha\beta}=G_{\alpha\beta}/(n_{\beta}+n_{\bar{\beta}})$.  Parameter space was explored through a weighted Markov Chain Monte Carlo trace. The likelihood was expressed in terms of $\chi^2$, the sum of the squared discrepancies for each bin of the three balance functions, $\chi^2\equiv\sum_{i} (y_i-y_i^{\rm(exp)})^2 / \sigma_i^2$, ignoring the bin for $\Delta y<0.1$. The likelihood was proportional to $e^{-\chi^2/2}$. The uncertainties for each point, $\sigma_i$, were assigned rather arbitrarily as 7.5\% of the absolute value of the  measured value plus a second contribution that fell from 0.01 to zero linearly as you increased $\Delta\eta$ to the maximum value for measurement. Since the calculation for each point required a few minutes of CPU time, we applied a model emulator described in \cite{Novak:2013bqa,Gomez:2012ak}. The emulator strategy involves interpolating data from a number of sampling runs, in this case 1024 runs. The emulator was compared to full-model runs to test its accuracy.

\begin{figure}
\centerline{\includegraphics[width=0.5\textwidth]{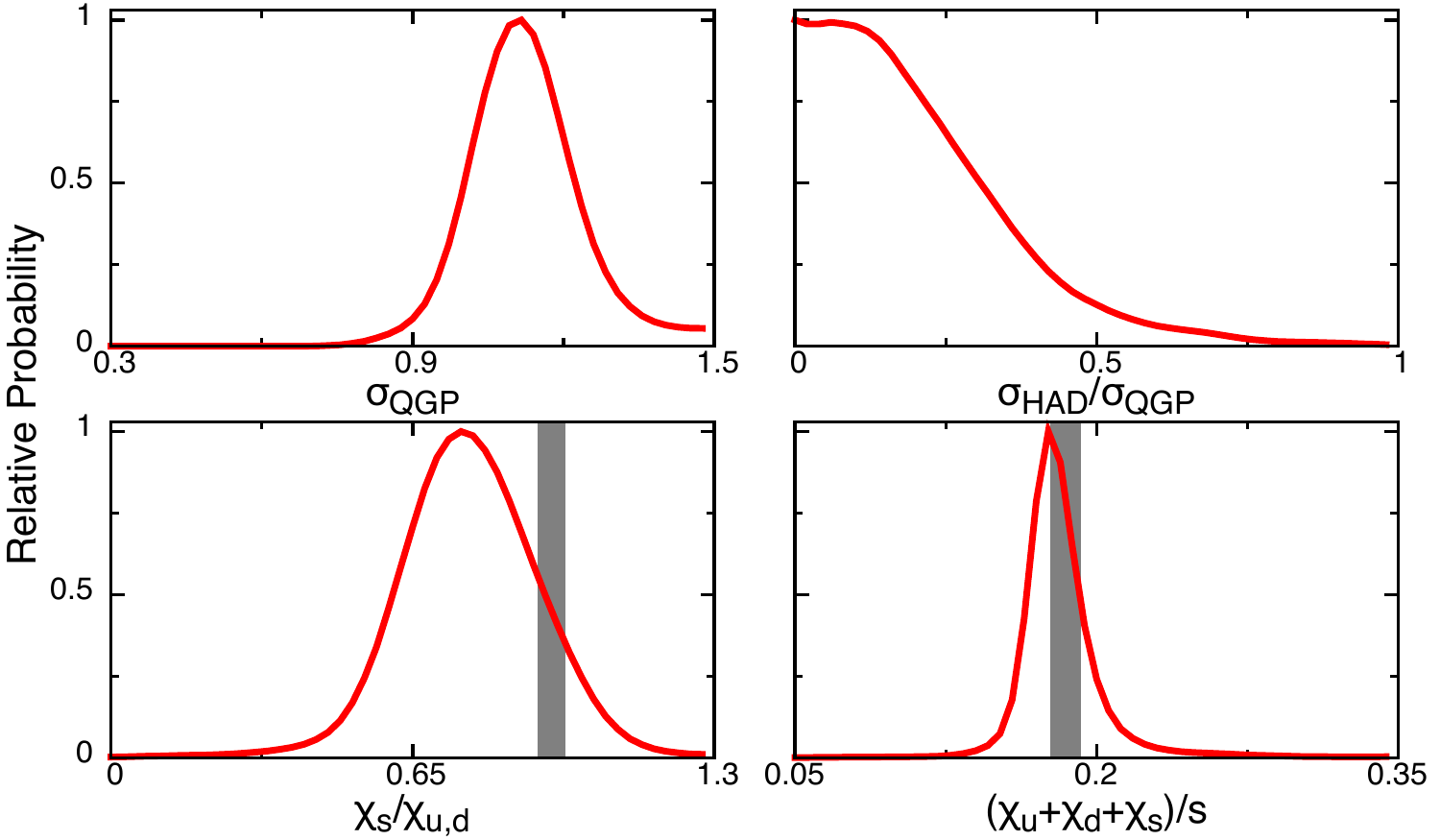}}
\caption{\label{fig:1dlikelihood}
Posterior distribution of parameters from MCMC trace.  The two gray lines show expectations of quark chemistry from lattice gauge theory.}
\end{figure}
The preferred region of parameter space is close to expectations. For charge created in the initial creation of a thermalized QGP, it appears that balancing charges separated on the order of one unit of spatial rapidity by the time of breakup. For the second wave of production, at or after hadronization, it would appear that the spread would be approximately a fourth of that.  The gray bands in Fig. \ref{fig:1dlikelihood} represent the range of $\chi_{ss}/s$ and $(\chi_{uu}+\chi_{dd}+\chi_{ss})/s$ from lattice calculations as the temperature varied from 250 to 350 MeV \cite{Ratti:2011au}. The preferred value for $\chi_{ss}/s$ seems to be 20\% below the lattice values whereas the sum $(\chi_{uu}+\chi_{dd}+\chi_{ss})/s$ is consistent with the lattice calculations .

\begin{figure}
\centerline{
\includegraphics[width=0.5\textwidth]{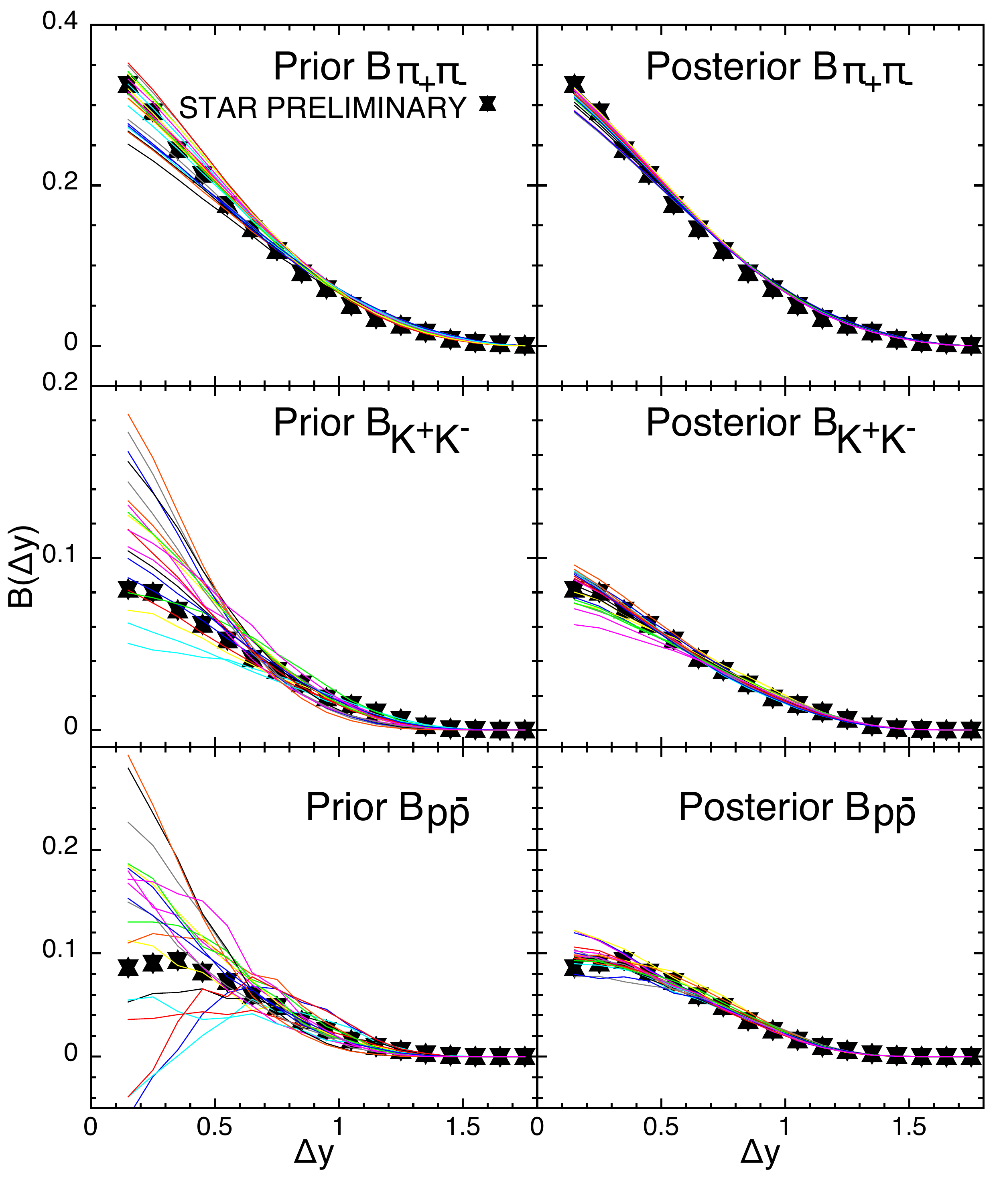}}
\caption{\label{fig:priorvpost}
Left panels: Balance functions calculated from 20 points drawn randomly from the original parameter space (prior). Right panels: Same as the left, but with 20 points taken randomly from the MCMC trace (posterior).  
}
\end{figure}
To validate the MCMC procedure and the emulator, 20 points in the four-dimensional parameter space were taken from the posterior distribution, i.e. weighted by the likelihood. The model was then re-run using these parameters. The resulting balance functions were then compared to the data and shown in Fig. \ref{fig:priorvpost}. The calculations appear quite successful in reproducing the experimental balance functions. In contrast 20 points were randomly drawn from the prior distribution. The resulting balance functions are also shown in Fig. \ref{fig:priorvpost}, and found to strongly differ from the experimental results. This illustrates how strongly balance functions are being constrained, and also shows that the $K^+K^-$ and $p\bar{p}$ balance functions represent most of the resolving power.

In summary, charge balance functions from calculations described in \cite{Pratt:2012dz} were compared to preliminary STAR results. The comparison led to a determination of the quark chemistry which was within 20\% of expectations based on assumptions of an isentropic expansion of a chemically equilibrated (according to lattice) QGP. The analysis also provides validation of the two-wave nature of quark production in relativistic heavy ion collisions. Given the schematic nature of the model, one could not have expected to reach quantitative conclusions better than the 20\% level. This study should serve as motivation for both improved and more detailed microscopic modeling, and for additional data, both at RHIC and at the LHC. Like femtoscopic two-particle correlations, charge balance correlations intrinsically carry six dimensions of information, and the comparison of balance functions for different species provide even richer and more constraining insight. Incoming data from STAR, ALICE, CMS and ATLAS should greatly improve our understanding of the chemical and diffusive properties of super-hadronic matter created in heavy ion collisions.

\begin{acknowledgments}
The authors are indebted to Hui Wang for providing preliminary STAR data, and to Gary Westfall for providing the efficiency routine to simulate STAR's acceptance. This work was supported by the National Science Foundation's Cyber-Enabled Discovery and Innovation Program through grant NSF-0941373 and by the Department of Energy Office of Science through grant number DE-FG02-03ER41259.
\end{acknowledgments}

\end{document}